\documentclass[sigconf, screen, nonacm]{acmart} 

\newif\ifanonymous
\anonymousfalse  %

\ifanonymous
  \newcommand{\ToolEval}{[EvalTool]}      %
  \newcommand{\ToolGen}{[GenTool]}        %
  \newcommand{\selfcite}[1][]{[anonymous]}  %
\else
  \newcommand{\ToolEval}{JudgeGPT}
  \newcommand{\ToolGen}{RogueGPT}

  \newcommand{\selfcite}[1][]{\cite{loth_blessing_2024,loth2026collateraleffects,loth2026eroding,loth2026roguegpt,loth2026verification,loth2026originlens}}
\fi

\copyrightyear{2026}
\acmYear{2026}
\setcopyright{cc}
\setcctype{by}
\acmConference[WWW Companion '26]{Companion Proceedings of the ACM Web Conference 2026}{April 13--17, 2026}{Dubai, United Arab Emirates}
\acmBooktitle{Companion Proceedings of the ACM Web Conference 2026 (WWW Companion '26), April 13--17, 2026, Dubai, United Arab Emirates}
\acmPrice{}
\acmDOI{10.1145/3774905.3795832}
\acmISBN{979-8-4007-2308-7/2026/04}

\usepackage{algorithmic}
\usepackage{graphicx}
\usepackage{subcaption}
\usepackage{siunitx}
\usepackage{enumitem}
\usepackage{microtype}
\usepackage{textcomp}
\usepackage{appendix}
\usepackage{hyperref}

\usepackage[hyphenbreaks]{breakurl}
\usepackage{xurl}
\def\BibTeX{{\rm B\kern-.05em{\sc i\kern-.025em b}\kern-.08em
    T\kern-.1667em\lower.7ex\hbox{E}\kern-.125emX}}
\usepackage{cleveref}
\usepackage{booktabs}
    
\usepackage{xcolor}

 \usepackage{pifont}%
\newcommand{\cmark}{\ding{51}}%
\newcommand{\xmark}{\ding{55}}%

\usepackage{listings}
\usepackage{fancyvrb}
\usepackage{framed}
\usepackage{color}

\usepackage{tabularx}
\definecolor{light-gray}{gray}{0.95}
\hypersetup{
   breaklinks=true,   %
   colorlinks=true    %
}

\usepackage{fancyhdr}
 
\begin{document}

\title[Eroding the Truth-Default]{Eroding the Truth-Default: A Causal Analysis of Human Susceptibility to Foundation Model Hallucinations and Disinformation in the Wild}
\titlenote{This research is funded by the European Union project CyberSecDome under the grant agreement 101120779. Published in WWW Companion '26, pp.~1125--1132. DOI: \href{https://doi.org/10.1145/3774905.3795832}{10.1145/3774905.3795832}.}

\author{Alexander Loth}
\email{alexander.loth@stud.fra-uas.de}
\orcid{0009-0003-9327-6865}
\affiliation{%
  \institution{Frankfurt University of Applied Sciences}
  \city{Frankfurt am Main}
  \country{Germany}
}

\author{Martin Kappes}
\email{kappes@fra-uas.de}
\orcid{0000-0002-8768-8359}
\affiliation{%
  \institution{Frankfurt University of Applied Sciences}
  \city{Frankfurt am Main}
  \country{Germany}
}

\author{Marc-Oliver Pahl}
\email{marc-oliver.pahl@imt-atlantique.fr}
\orcid{0000-0001-5241-3809}
\affiliation{%
  \institution{IMT Atlantique, UMR IRISA, Chaire Cyber CNI}
  \city{Rennes}
  \country{France}
}

\begin{abstract}
As foundation models (FMs) approach human-level fluency, distinguishing synthetic from organic content has become a key challenge for Trustworthy Web Intelligence. 

This paper presents \textbf{JudgeGPT} and \textbf{RogueGPT}, a dual-axis framework that decouples ``authenticity'' from ``attribution'' to investigate the mechanisms of human susceptibility. 
Analyzing 918 evaluations across five FMs (including GPT-4 and Llama-2), we employ Structural Causal Models (SCMs) as a principal framework for formulating testable causal hypotheses about detection accuracy. 

Contrary to partisan narratives, we find that political orientation shows a negligible association with detection performance ($r=-0.10$). Instead, ``fake news familiarity'' emerges as a candidate mediator ($r=0.35$), suggesting that exposure may function as adversarial training for human discriminators. 
We identify a ``fluency trap'' where GPT-4 outputs (HumanMachineScore: 0.20) bypass Source Monitoring mechanisms, rendering them indistinguishable from human text. 

These findings suggest that ``pre-bunking'' interventions should target cognitive source monitoring rather than demographic segmentation to ensure trustworthy information ecosystems.
\end{abstract}



\maketitle

\pagestyle{fancy}
\fancyhf{}
\fancyhead[L]{\textit{A.\ Loth, M.\ Kappes, M.-O.\ Pahl}}
\fancyhead[R]{\textit{Published in WWW Companion '26, pp.\ 1125--1132}}
\fancyfoot[C]{\thepage}
\renewcommand{\headrulewidth}{0pt}

\section{Introduction}
\label{sec:introduction}

The rapid proliferation of foundation models, particularly large language models (LLMs), presents both opportunities and challenges for web intelligence. Han et al.~\cite{han2025trustworthy} define trustworthy machine learning as encompassing robustness, privacy, and causal integrity---properties that extend beyond model performance to the entire data-model-human pipeline. As these models improve, distinguishing between human-written and machine-generated content has become a central challenge for maintaining trustworthy information ecosystems online~\cite{loth2024blessing}. Recent work has examined the causal reasoning capabilities of LLMs~\cite{chi2025unveiling} and their tendency to hallucinate~\cite{li2024llmhallucinate}, yet how users perceive and evaluate foundation model outputs ``in the wild'' remains underexplored.

Building trustworthy AI systems requires understanding not only the technical reliability of models but also the causal factors that influence human susceptibility to AI-generated misinformation and disinformation~\cite{loth2021decisively}. Yao et al.~\cite{yao2025trustworthy} frame web data as inherently ``imperfect,'' subject to noise and annotation errors; we extend this insight by treating human perception errors as a form of imperfect annotation that reveals latent variables governing trust. Following Liu et al.~\cite{liu2025causal}, we approach the analysis of human judgment patterns as a method of causal discovery---identifying the hidden factors that mediate susceptibility.

To address this gap, we developed \textbf{\ToolEval} (\url{https://github.com/aloth/JudgeGPT}), an interactive, open-source platform for systematically studying human perception of AI-generated content, and \textbf{\ToolGen} (\url{https://github.com/aloth/RogueGPT}), a complementary content generation engine that produces diverse news fragments using multiple foundation models. Together, these tools enable investigation of causal relationships between user characteristics, content properties, and judgment accuracy.

Our research addresses the following questions with implications for trustworthy foundation models:

\begin{itemize}[leftmargin=*]
    \item \textbf{Causal Factors in Susceptibility:} What demographic and experiential factors causally influence users' ability to distinguish foundation model outputs from human-written content?
    \item \textbf{Model-Specific Vulnerabilities:} Do different foundation models vary in their ability to produce human-like content, and what are the implications for trustworthy AI deployment?
    \item \textbf{Learning and Adaptation:} Can exposure to AI-generated content improve detection abilities over time, suggesting potential for ``pre-bunking'' interventions?
\end{itemize}

\subsection{Contributions}
\label{subsec:contributions}

This paper offers three contributions:

\begin{itemize}[leftmargin=*]
    \item \textbf{\ToolEval\ and \ToolGen\ Platforms:} Open-source tools for continuous data collection on human perception of foundation model outputs, featuring dual-axis assessment (authenticity and source attribution) with multilingual support.
    \item \textbf{Causal Analysis of Human Perception:} Empirical findings from 154 participants and 918 evaluations revealing how fake news familiarity, political orientation, and other factors causally influence detection accuracy.
    \item \textbf{Model Comparison Insights:} Evidence that GPT-4 produces more human-like text than other models tested, with implications for trustworthy foundation model development and deployment.
\end{itemize}

\subsection{Paper Structure}
\label{subsec:structure}

Section~\ref{sec:related} situates our work within related research. Section~\ref{sec:judgegpt} describes the \ToolEval\ and \ToolGen\ platforms. Section~\ref{sec:results} reports our primary findings. Section~\ref{sec:conclusion} discusses implications for trustworthy foundation models and future research directions.

\section{Related Work}
\label{sec:related}

Research on trustworthy foundation models intersects with several established areas that inform our approach.

\subsection{Trustworthy AI and Foundation Models}

The challenge of building trustworthy AI systems has received attention as foundation models become widespread~\cite{han2025trustworthy}. Weidinger et al.~\cite{weidinger2022taxonomy} provide a taxonomy of LLM risks including misinformation generation and human overreliance on fluent outputs. Key concerns include model reliability under noisy or adversarial conditions~\cite{han2018coteaching}, the causal reasoning capabilities of LLMs~\cite{chi2025unveiling}, and understanding why models hallucinate~\cite{li2024llmhallucinate}. While Chi et al.~\cite{chi2025unveiling} demonstrate that LLMs lack ``Level 2'' causal reasoning, our data reveals that humans falsely attribute this capability to them due to output fluency---a phenomenon Bender et al.~\cite{bender2021stochastic} term ``stochastic parrots,'' where coherent-sounding text masks the absence of genuine understanding. These technical dimensions of trustworthiness are complemented by the need to understand human factors---specifically, how users perceive and interact with AI-generated content in web environments~\cite{yao2025trustworthy}.

\subsection{Cognitive Frameworks for AI Perception}

Our analysis draws on established cognitive science frameworks. The Source Monitoring Framework~\cite{johnson1993source} explains how humans use heuristic cues (detail, fluency) rather than verified source information when attributing text origins---we term this the ``Fluency Trap.'' Truth-Default Theory, recently applied to AI~\cite{markowitz2024generative}, suggests that AI systems exhibit truth bias, which may create ``validity illusions'' for users who assume AI outputs are accurate; fluent foundation model outputs may thus be \textit{eroding the truth-default} that humans apply to textual communication. Finally, dual-process theory~\cite{kahneman2011thinking} explains why detection accuracy degrades over time: sustained analytical effort (System 2) depletes cognitive resources, leading participants to rely on faster but less accurate heuristics.

\subsection{Human Perception and Fake News Detection}

The intersection of generative AI and misinformation presents a dual challenge: while LLMs can assist in detection, they also lower barriers to fake news generation~\cite{loth2024blessing,loth2026collateraleffects}. Automated fake news detection has employed techniques from multimodal architectures~\cite{wang_yaqingwangeann-kdd18_2018} to transformer-based systems like GROVER~\cite{Grover}. The RAID benchmark~\cite{dugan2024raid} provides the current standard for evaluating machine-generated text detectors, demonstrating that automated systems struggle with adversarial attacks and domain shifts. Our work complements RAID by providing \textit{human-centric} metrics: while RAID measures algorithmic detection, \ToolEval\ measures human detection---both are necessary for a complete picture of trustworthiness in web intelligence.

Studies have investigated how demographic factors influence credibility judgments~\cite{yang_others_2021, bradshaw_industrialized_2021, van2022misinformation} and factors influencing trust in machine-generated content~\cite{huschens2023trustchatgptperceived}. Clark et al.~\cite{clark2021humans} found that human evaluation of generated text is unreliable, with evaluators often unable to distinguish GPT-3 outputs from human writing. Ippolito et al.~\cite{ippolito2020automatic} demonstrated that automatic detection is easiest precisely when humans are fooled, suggesting complementary roles for algorithmic and human detection. The MIST test~\cite{maertens2024mist} measures misinformation susceptibility using psychometric analysis. However, most work focuses on binary classification (real vs. fake), missing the nuanced spectrum of human perception.

\subsection{Pre-bunking and Psychological Inoculation}

Our finding on learning effects connects to the literature on psychological inoculation against misinformation. Roozenbeek et al.~\cite{roozenbeek2022psychological, roozenbeek2024prebunking} have demonstrated that gamified ``pre-bunking'' interventions can improve resilience against misinformation on social media. The Debunking Handbook~\cite{lewandowsky2020debunking} synthesizes evidence on effective correction strategies, emphasizing that prebunking is more effective than debunking. Google Jigsaw's collaboration on EU election pre-bunking campaigns~\cite{jigsaw2024prebunking} demonstrates real-world applications of this research for web intelligence. We position \ToolEval\ as a \textit{dynamic pre-bunking environment}: beyond passive measurement, the platform's iterative exposure to AI-generated content may itself serve as a form of cognitive inoculation.

Table~\ref{tab:tool_comparison_compact} positions \ToolEval\ relative to existing approaches.

\begin{table}[htbp]
    \centering
    \small
    \begin{tabular}{|l|c|c|c|}
        \hline
        \textbf{Tool} & \textbf{Human Eval} & \textbf{Dual-Axis} & \textbf{Multi-LLM} \\ \hline
        GROVER~\cite{Grover} & \xmark & \xmark & \xmark \\ \hline
        MIST~\cite{maertens2024mist} & \cmark & \xmark & \xmark \\ \hline
        FakeGPT~\cite{huang2024fakegpt} & \xmark & \xmark & \xmark \\ \hline
        \textbf{\ToolEval\ (Ours)} & \cmark & \cmark & \cmark \\
        \hline
    \end{tabular}
    \caption{\ToolEval\ provides dual-axis human evaluation across multiple foundation models.}
    \label{tab:tool_comparison_compact}
\end{table}

\ToolEval\ extends prior work by: (1) capturing nuanced perceptions using continuous scales for both authenticity and source attribution; (2) enabling demographic analysis to identify causal factors in AI content perception; (3) comparing multiple foundation models to assess relative ``human-likeness''; and (4) supporting multilingual evaluation for cross-cultural insights.

\section{\ToolEval\ and \ToolGen\ Platforms}
\label{sec:judgegpt}

Our dual-system framework combines \textbf{\ToolEval} (\url{https://github.com/aloth/JudgeGPT}), an interactive platform for evaluating human perception of content authenticity and source, with \textbf{\ToolGen} (\url{https://github.com/aloth/RogueGPT}), a generative engine producing news fragments using multiple foundation models. Together, these platforms form the empirical foundation for our ongoing research program on human-AI interaction in the context of misinformation~\cite{loth2024blessing,loth2026collateraleffects}.

\subsection{\ToolEval: Human Evaluation Platform}

\ToolEval\ implements a Streamlit-based web interface that presents news fragments and captures user evaluations along two continuous dimensions:
\begin{itemize}[leftmargin=*]
    \item \textbf{Perceived Authenticity (LegitFakeScore):} From ``Completely Fake'' to ``Completely Legitimate'' --- capturing how believable the content appears
    \item \textbf{Source Attribution (HumanMachineScore):} From ``Definitely Human-Written'' to ``Definitely Machine-Generated'' --- capturing perceived authorship
\end{itemize}

This dual-axis approach enables analysis of whether humans can distinguish AI-generated content (source attribution) independently of whether they find it credible (authenticity assessment). The platform supports four languages (English, German, French, Spanish) with automatic browser language detection. After obtaining informed consent, participants provide demographic information (age, gender, education, political orientation, fake news familiarity) before evaluating news fragments. The system logs response times and interaction patterns alongside judgments, enabling temporal analysis of learning and fatigue effects.

\subsection{\ToolGen: Content Generation Engine}

\ToolGen\ generates news fragments using multiple foundation models, enabling comparative analysis of how different models vary in producing human-like content. The currently supported models include:
\begin{itemize}[leftmargin=*]
    \item \textbf{OpenAI:} GPT-4, GPT-4o, GPT-3.5-turbo
    \item \textbf{Meta:} Llama-2-13b-chat
    \item \textbf{Google:} Gemma-7b-it
    \item \textbf{Mistral AI:} Mistral-7b-instruct
    \item \textbf{Microsoft:} Phi-3-mini-4k-instruct
\end{itemize}

This multi-model approach informs trustworthy AI deployment: if certain models consistently evade human detection, they pose greater risks for misinformation propagation on the web.

\subsection{Data Collection and Metrics}

The platform collects three types of data for causal analysis:
\begin{enumerate}[leftmargin=*]
    \item \textbf{Participant demographics:} Age, gender, education level, political orientation, fake news familiarity, newspaper subscriptions, and geographic location --- enabling analysis of susceptibility factors
    \item \textbf{Fragment metadata:} Source (human or AI), topic, language, generating model, and veracity (based on factual accuracy) --- enabling model comparison
    \item \textbf{Assessment metrics:} Continuous slider positions for both dimensions, response times, and interaction patterns --- enabling behavioral analysis
\end{enumerate}

As of December 2025, the dataset includes 154 validated participants and 918 evaluations across 2301 news fragments in four languages.

\section{Findings}
\label{sec:results}

This section presents findings from 918 valid participant responses, examining factors associated with human perception of foundation model outputs and comparative model performance.

\subsection{From Correlation to Causal Hypotheses via SCMs}
\label{subsec:scm}

While correlational analysis provides initial insights, we frame our findings within a Structural Causal Model (SCM)~\cite{pearl2009causality, imbens2015causal} to formulate explicit, testable causal hypotheses. We propose a Directed Acyclic Graph (DAG) with the following structure:

\begin{itemize}[leftmargin=*]
    \item $T$ \textbf{(Treatment):} Model Architecture (e.g., GPT-4 vs. Mistral vs. Llama-2~\cite{meta2024llama3})
    \item $M$ \textbf{(Mediator):} Textual Features---decomposed into three interpretable dimensions: (1) \textit{Surface Fluency}: sentence length variance, type-token ratio (lexical diversity), grammatical error rate; (2) \textit{Discourse Coherence}: connective frequency (``however,'' ``therefore''), referential consistency, topic continuity; and (3) \textit{Causal Consistency}: logical entailment patterns, temporal plausibility, factual grounding. While we do not computationally extract these features in the current study, prior work~\cite{ippolito2020automatic, clark2021humans} demonstrates their differential role in human detectability---with surface fluency being the most salient cue and causal consistency the least utilized.
    \item $C$ \textbf{(Confounders):} User Demographics (fake news familiarity, age, education, political orientation)
    \item $Y$ \textbf{(Outcome):} Source Attribution (HumanMachineScore)
\end{itemize}

This formalization enables claims using Pearl's \textit{do}-calculus: ``Controlling for political orientation ($C$), the direct causal effect of Model Architecture ($T$) on detection failure ($Y$) is mediated by Perceived Fluency ($M$).'' Formally, $P(Y|do(T))$ captures the interventional effect of deploying different foundation models on human susceptibility. This language aligns with the causal inference focus of the workshop~\cite{han2025trustworthy, liu2025causal}.

\textit{Methodological Note:} We acknowledge that our observational design cannot fully establish causality in the interventional sense. While SCMs provide a principled framework for reasoning about causal relationships, the correlations we report should be interpreted as \textit{causal hypotheses} requiring validation through randomized experiments (e.g., manipulating fake news exposure or model assignment). Our DAG represents a testable causal model rather than confirmed causal knowledge. Critically, we assume no unmeasured confounders between perceived fluency ($M$) and susceptibility ($Y$)---yet factors such as cognitive load, reading speed, or domain expertise could confound this relationship. Future work should employ sensitivity analyses~\cite{imbens2015causal} to quantify how robust our estimates are to violations of this assumption, and experimental designs that directly manipulate textual features while holding content constant.

\subsection{Associational Patterns: Fake News Familiarity and Political Orientation}
\label{subsec:findings_poli_familiarity}

\begin{figure}[!ht]
    \centering
    \includegraphics[width=0.45\textwidth]{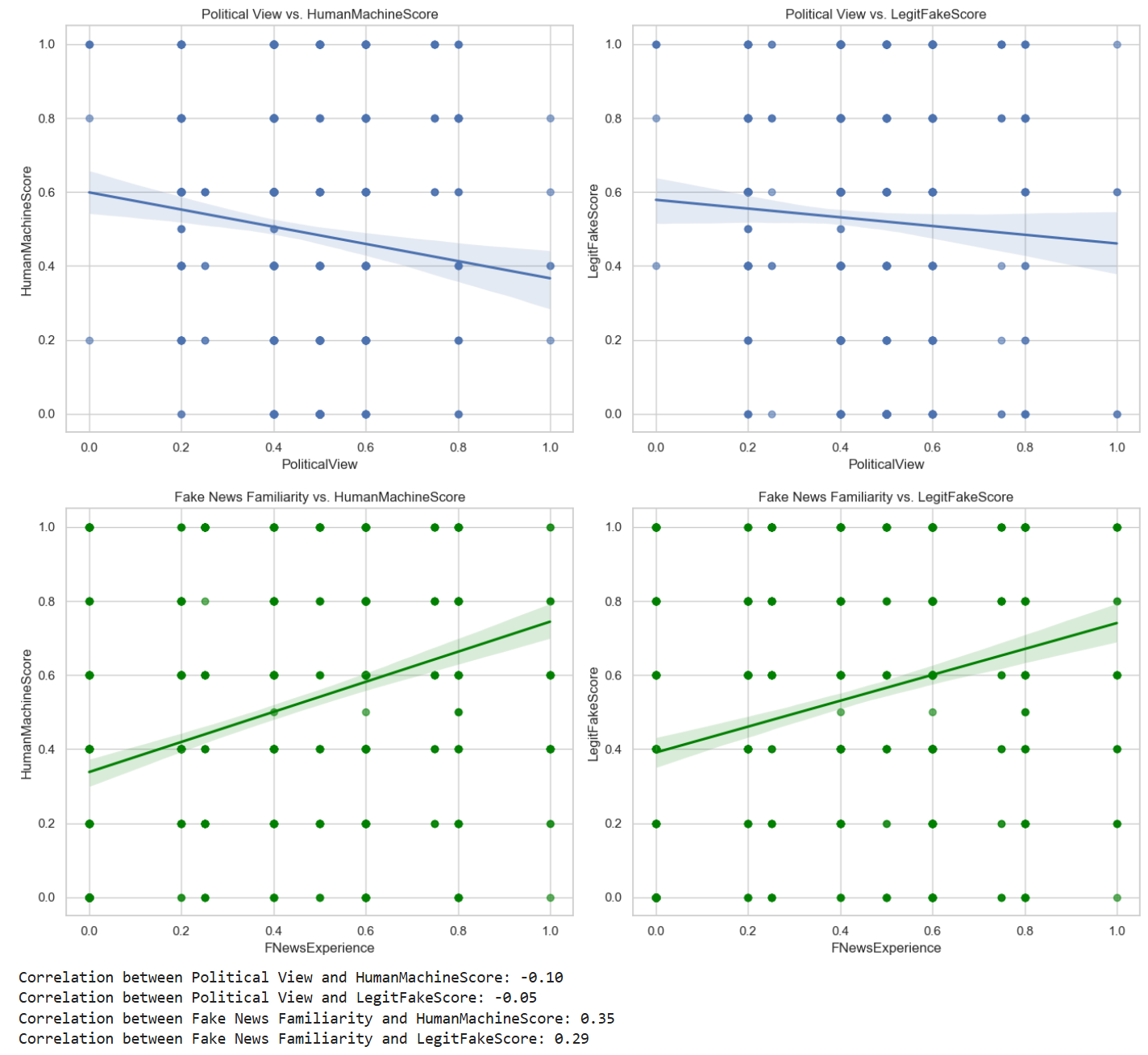}
    \Description{Heatmap of correlation coefficients between participant characteristics (e.g., fake news familiarity and political orientation) and the two judgment metrics (source attribution and legitimacy).}
    \caption{Correlations between participant characteristics and judgment scores. Fake news familiarity shows stronger predictive power than political orientation.}
    \label{fig:figure1}
\end{figure}

We analyzed associations between participant characteristics (confounders $C$) and judgment accuracy (Figure~\ref{fig:figure1}):

\begin{itemize}[leftmargin=*]
    \item \textbf{Fake news familiarity is associated with accuracy:} Moderate positive correlations exist between fake news familiarity and both source attribution ($r=0.35$) and legitimacy assessment ($r=0.29$). Under our proposed DAG, this association is consistent with prior misinformation experience improving detection capabilities---a hypothesis supported by findings from our larger reproduction study~\cite{loth2026verification}.
    \item \textbf{Political orientation shows weak associations:} Only weak negative correlations exist between political extremity and judgment accuracy ($r=-0.10$ for source, $r=-0.05$ for legitimacy), suggesting political orientation is not a primary factor---consistent with Pennycook and Rand's~\cite{pennycook2019lazy} finding that analytical thinking, not partisan bias, predicts fake news susceptibility.
\end{itemize}

These associational patterns, interpreted through our causal framework, suggest that interventions targeting fake news awareness may be more effective than those targeting political groups---a hypothesis warranting experimental validation.

\subsection{Foundation Model Comparison: Perceived Humanness}
\label{subsec:findings_model_perf}

A notable finding for trustworthy foundation models is the variation in how ``human-like'' different models appear to users. Table~\ref{tab:model_performance} shows the average HumanMachineScore by model (lower = more human-like).

\begin{figure*}[!ht]
    \centering
    \includegraphics[width=0.9\textwidth]{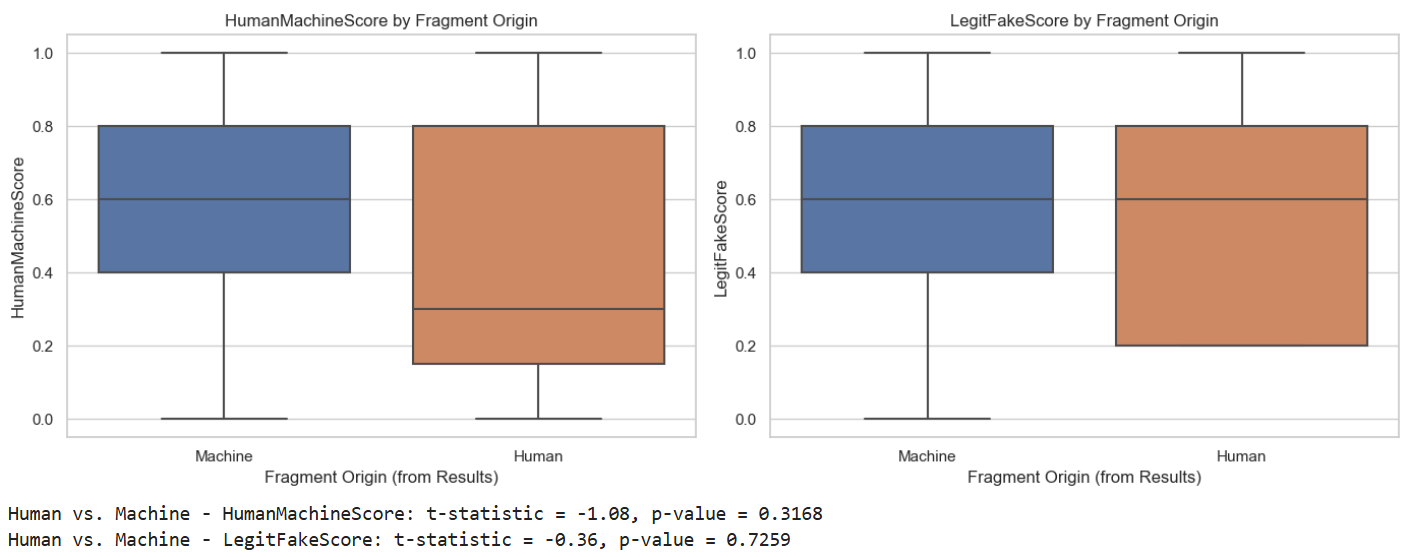}
    \Description{Side-by-side boxplots comparing the distributions of HumanMachineScore and LegitFakeScore for fragments by actual source (human-written vs. AI-generated).}
    \caption{Comparison of HumanMachineScore and LegitFakeScore for fragments by actual source (Human vs. AI). Boxplots show median and interquartile ranges. Independent $t$-tests found no significant mean differences between human-written and AI-generated groups for either metric, highlighting the challenge of distinguishing foundation model outputs.}
    \label{fig:figure2}
\end{figure*}

\textbf{Result:} GPT-4 outputs received lower scores (0.20) than all other models (next best: 0.44), indicating GPT-4 produces text that participants perceive as more human-like. This finding has implications for foundation model trustworthiness: as models become more capable, human detection becomes less reliable. Strikingly, when comparing human-written and AI-generated fragments in aggregate (Figure~\ref{fig:figure2}), independent $t$-tests found no significant mean differences between groups for either HumanMachineScore or LegitFakeScore---participants could not reliably distinguish AI content from human content at the population level.

\begin{table}[htbp]
    \centering
    \caption{Average HumanMachineScore by foundation model (lower = more human-like).}
    \label{tab:model_performance}
    \small
    \begin{tabular}{|l|c|}
    \hline
    \textbf{Model} & \textbf{Avg. Score} \\ \hline
    \texttt{openai\_gpt-4} & 0.200 \\ \hline
    \texttt{mistralai\_mistral-7b} & 0.442 \\ \hline
    \texttt{openai\_gpt-4o} & 0.468 \\ \hline
    \texttt{google\_gemma-7b} & 0.499 \\ \hline
    \texttt{meta\_llama-2-13b} & 0.526 \\ \hline
    \texttt{openai\_gpt-35-turbo} & 0.548 \\ \hline
    \texttt{microsoft\_Phi-3-mini} & 0.580 \\ \hline
    \end{tabular}
\end{table}

\subsection{Learning Effects and Behavioral Patterns}
\label{subsec:findings_learning}

We examined whether participants' judgment criteria changed over the course of the experiment, possibly due to learning or fatigue. Figure~\ref{fig:figure4} plots the average scores as a function of response order.

\paragraph{Learning Effects:} The HumanMachineScore shows a downward trend early in sessions before stabilizing (Figure~\ref{fig:figure4}), suggesting participants improve at detecting AI text with exposure. This aligns with research on psychological inoculation against misinformation~\cite{roozenbeek2022psychological}. Our larger-scale analysis reveals asymmetric cognitive fatigue: fake news detection accuracy degrades under sustained exposure while AI source detection remains relatively stable~\cite{loth2026collateraleffects}. This fatigue effect is consistent with dual-process theory~\cite{kahneman2011thinking}, where sustained System 2 effort leads to resource depletion. This learning effect suggests that targeted ``pre-bunking'' training could leverage natural learning mechanisms to improve detection capabilities.

\begin{figure*}[!ht]
    \centering
    \includegraphics[width=0.875\textwidth]{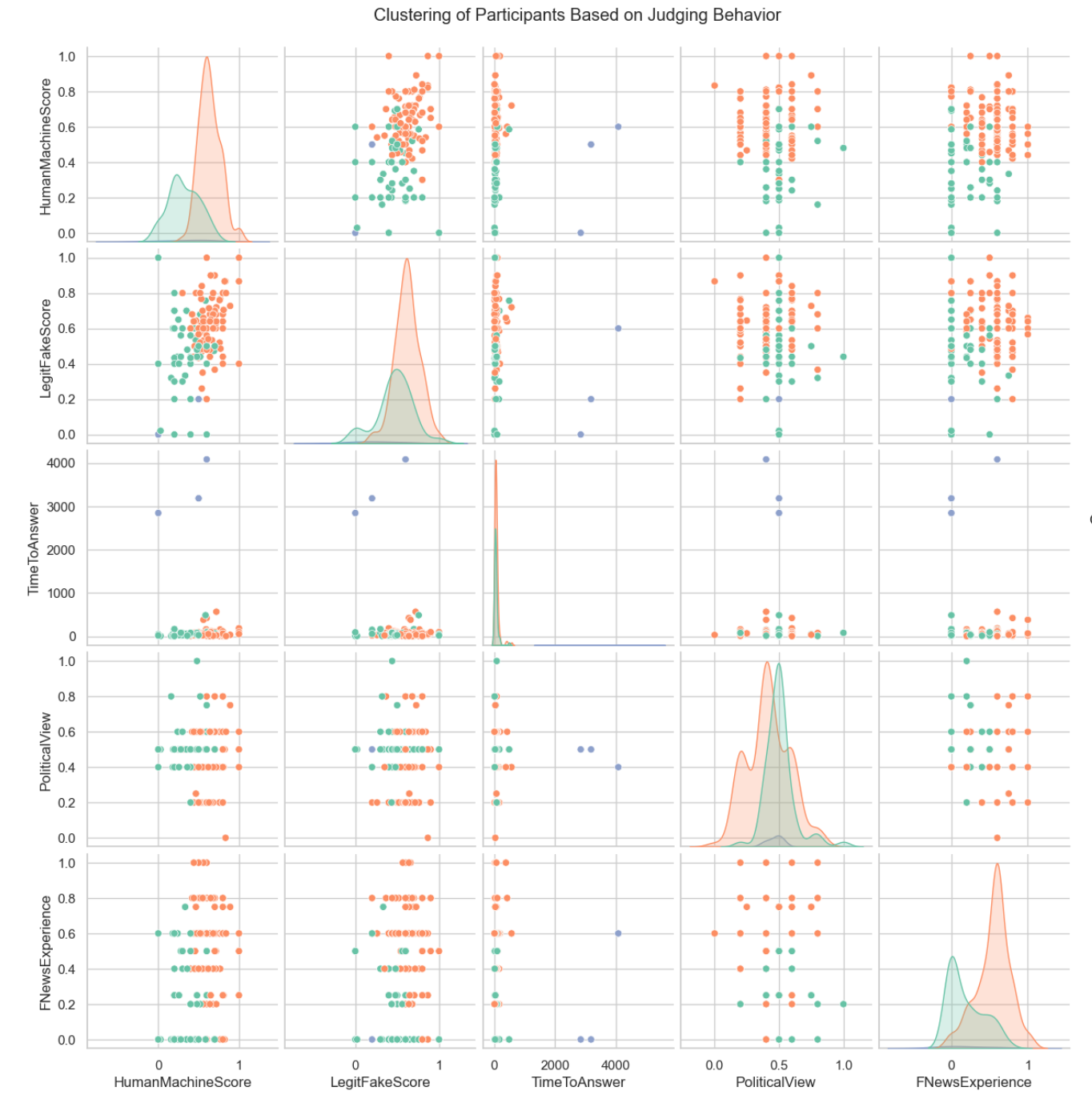}
    \Description{Pair-plot showing relationships among participant-level variables (HumanMachineScore, LegitFakeScore, TimeToAnswer, PoliticalView, and FNewsExperience), with points colored by inferred judging-behavior cluster.}
    \caption{Pair-plot visualization of participant judging behavior clusters. Each point represents a participant, colored by cluster. The plots show relationships between HumanMachineScore, LegitFakeScore, TimeToAnswer, PoliticalView, and FNewsExperience. Distinct clusters indicate different judgment behavior patterns that may be causally linked to participant characteristics.}
    \label{fig:figure3}
\end{figure*}

\begin{figure*}[!ht]
    \centering
    \includegraphics[width=0.9\textwidth]{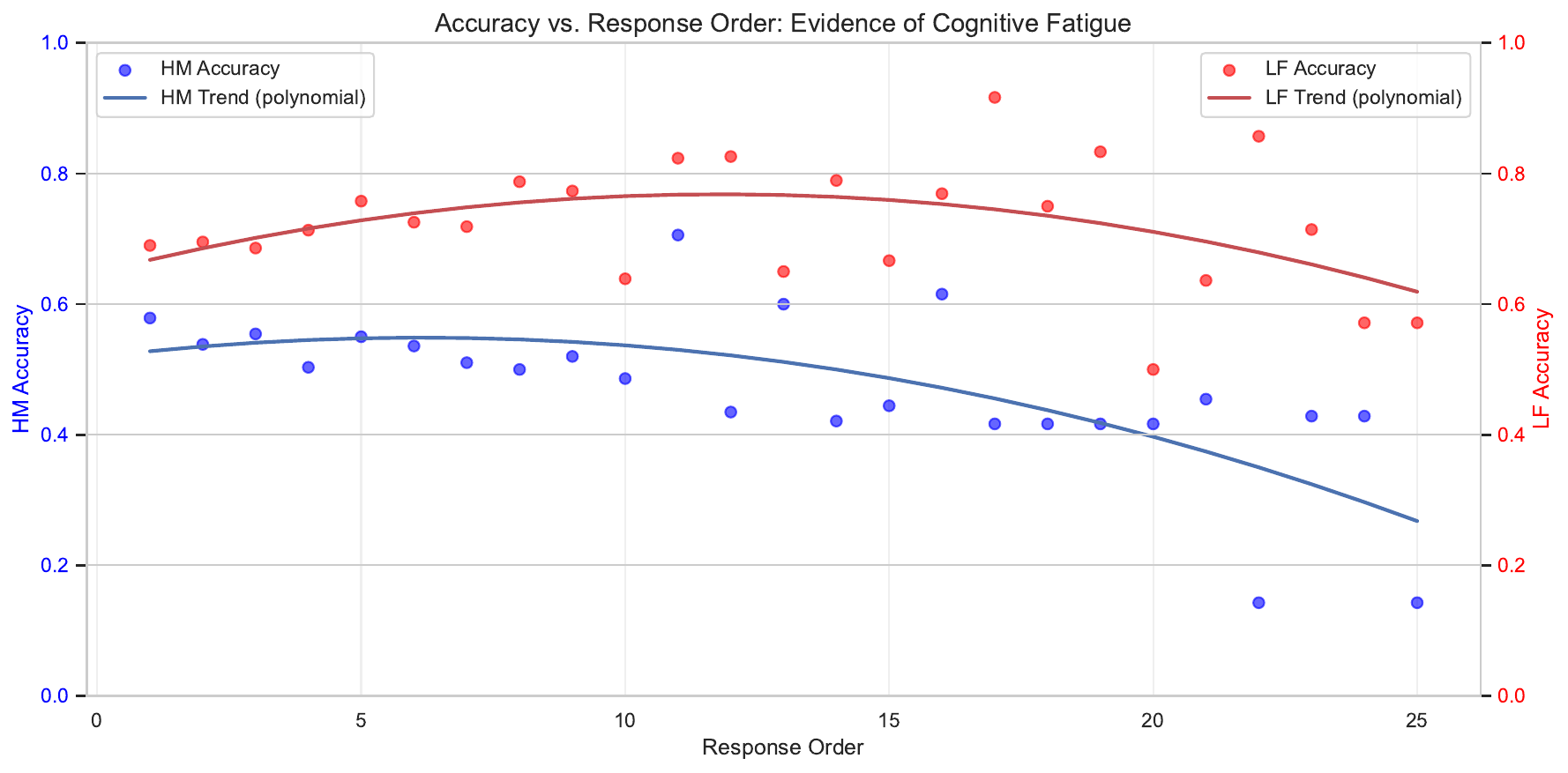}
    \Description{Time-series plot of average judgment scores versus response order, illustrating an early decrease in HumanMachineScore (learning) and a later drop in LegitFakeScore (possible fatigue).}
    \caption{Judgment Scores vs. Response Order. HumanMachineScore and LegitFakeScore are plotted over the sequence of trials for each participant (with smoothing). The HumanMachineScore tends to decrease initially (learning effect), while the LegitFakeScore drops later in the session (possible fatigue or strategy shift toward increased skepticism).}
    \label{fig:figure4}
\end{figure*}

\paragraph{Behavioral Clustering:} Clustering analysis (Figure~\ref{fig:figure3}) revealed distinct ``judging profiles'' among participants:
\begin{itemize}[leftmargin=*]
    \item \textbf{Skeptical judges:} Consistently rated content as machine-generated and fake
    \item \textbf{Trusting judges:} Tended to rate content as human-written and legitimate---consistent with Truth-Default Theory applied to AI~\cite{markowitz2024generative}
    \item \textbf{Variable judges:} Showed inconsistent patterns, possibly reflecting genuine uncertainty
\end{itemize}

These profiles were influenced by factors including response time patterns and fake news familiarity, suggesting that individual differences in cognitive style affect susceptibility to AI-generated misinformation.

\paragraph{Heterogeneity of Treatment Effect (HTE):} Following Yao et al.~\cite{yao2025trustworthy}, we embrace the ``imperfect data'' inherent to web intelligence. The variation in human response times and the distinct ``Skeptical vs. Trusting'' clusters (Figure~\ref{fig:figure3}) are not merely noise---they represent \textit{heterogeneity of treatment effect} (HTE). Different users exhibit differential susceptibility to different models: Trusting judges are particularly vulnerable to GPT-4's fluency, while Skeptical judges show relative immunity. Identifying \textit{which users} are susceptible to \textit{which models} constitutes a causal discovery task~\cite{liu2025causal}. \ToolEval\ provides the granular, user-level data necessary to estimate these heterogeneous effects---a prerequisite for \textit{personalized AI safety interventions} that target specific cognitive profiles rather than broad demographic categories.

\section{Conclusion}
\label{sec:conclusion}

This research addresses the human element of Web Intelligence. Through our \ToolEval\ platform, we collected 918 human evaluations across four languages and identified the specific cognitive and demographic factors that mediate susceptibility to AI-generated misinformation and disinformation.

Our analysis yielded three findings:
\begin{enumerate}[leftmargin=*]
    \item \textbf{Fake news familiarity associated with detection accuracy:} Prior experience with misinformation is associated with detection accuracy ($r=0.35$ for source attribution, $r=0.29$ for legitimacy), while political orientation shows a negligible association ($r=-0.10$). This suggests that media literacy interventions may be more effective than demographic targeting.
    \item \textbf{The GPT-4 ``humanness gap'':} GPT-4 produces text perceived as more human-like than other models (HumanMachineScore: 0.20 vs. 0.44+), quantifying the challenge for trustworthy web intelligence. This gap has implications for AI governance and deployment policies.
    \item \textbf{Learning effects and cognitive inoculation:} Participants exhibit learning over evaluation sessions, with distinct judgment profiles emerging. This aligns with psychological inoculation research~\cite{roozenbeek2022psychological} and suggests that ``pre-bunking'' training could leverage natural learning mechanisms to improve detection capabilities.
\end{enumerate}

\textbf{Toward Cognitive Inoculation:} We align with the call by Han et al.~\cite{han2025trustworthy} to view trustworthiness as a property spanning data, model, and human interaction. Our findings support a shift from \textit{debunking} (post-hoc fact-checking) to \textit{pre-bunking} (proactive inoculation).

Inoculation theory~\cite{roozenbeek2022psychological, roozenbeek2024prebunking} posits that exposing individuals to a weakened form of a persuasive argument (the ``virus'') triggers the production of counter-arguments (the ``antibodies''), conferring resistance to future, stronger attacks. We position \ToolEval\ as a \textit{digital vaccine}: by exposing users to AI-generated text and providing immediate feedback (the ``weakened dose''), we force users to update their Source Monitoring heuristics~\cite{johnson1993source}. The learning curve observed in Figure~\ref{fig:figure4} visualizes this ``antibody production''---the strengthening of cognitive defenses through controlled exposure.

\textbf{Policy Recommendations for Web Intelligence:} Based on our findings, we propose three complementary interventions:

\begin{enumerate}[leftmargin=*]
    \item \textbf{Mandatory Provenance (C2PA):} Since human detection of GPT-4 is failing (HumanMachineScore: 0.20), relying on ``user vigilance'' is a failed policy. We advocate for cryptographic watermarking standards (e.g., C2PA)~\cite{loth2026originlens} to provide an external verification layer that does not depend on fallible human perception.
    \item \textbf{Cognitive Security Education:} Digital literacy programs must pivot from ``checking sources'' (increasingly impossible with AI-generated content) to ``checking logic'' (causal consistency)~\cite{loth2021decisively}. Users must be trained to identify the \textit{absence} of causal reasoning~\cite{chi2025unveiling} in hallucinations---a skill that targets the fluency trap directly.
    \item \textbf{Adversarial User Training:} Platforms should deploy adversarial training for users---periodically presenting synthetic content with verification prompts---to keep Source Monitoring heuristics calibrated to evolving model capabilities. Inspired by Google Jigsaw's EU election pre-bunking campaigns~\cite{jigsaw2024prebunking}, such interventions transform passive consumption into active cognitive exercise.
\end{enumerate}

\textbf{Toward Model-Side Transparency:} While our findings focus on human susceptibility, we identify concrete opportunities for operationalizing insights into model-level constraints and training objectives:

\begin{itemize}[leftmargin=*]
    \item \textbf{Stylistic Fingerprinting:} Models could be trained with an auxiliary loss that maintains detectable stylistic signatures (e.g., characteristic n-gram distributions, punctuation patterns) while preserving task performance. This creates a ``watermark by design'' that automated detectors can exploit.
    \item \textbf{Uncertainty Surfacing:} Rather than producing uniformly confident prose, models should be trained with calibration objectives that surface epistemic uncertainty through hedging language (``evidence suggests,'' ``it is unclear whether'') when generating claims with low confidence scores from the model's own probability distributions.
    \item \textbf{Provenance Metadata:} API responses should include machine-readable provenance headers (model ID, generation timestamp, temperature settings) enabling downstream verification pipelines~\cite{loth2026originlens}. This can be enforced at the infrastructure level without modifying model weights.
    \item \textbf{RLHF for Detectability:} Reinforcement learning from human feedback could incorporate a multi-objective reward: $R = \alpha R_{\text{utility}} + \beta R_{\text{detectability}}$, where the detectability reward is derived from human evaluator accuracy on held-out samples. This explicitly trades off deceptive capability against task performance.
    \item \textbf{Fluency-Gated Safety Classifiers:} Our finding that GPT-4's fluency bypasses Source Monitoring suggests that safety classifiers should weight fluency as a \textit{risk factor}---highly fluent outputs warrant additional scrutiny for factual grounding before release.
\end{itemize}

\textbf{Limitations and Future Work:} We acknowledge several constraints that bound our conclusions:

\begin{itemize}[leftmargin=*]
    \item \textbf{Sample Size and Regional Bias:} The current dataset ($N=154$ participants, 918 evaluations), while multilingual, is skewed toward educated European participants. This limits generalizability to populations with different media ecosystems, education levels, or cultural orientations toward authority and technology. We are actively expanding data collection with more diverse populations~\cite{loth2026verification}.
    \item \textbf{Content Domain Variation:} Our corpus spans political and general news topics but does not systematically analyze domain-specific effects. Detection accuracy may vary significantly between political misinformation, health claims, and scientific content---each domain carries distinct prior expectations and verification heuristics.
    \item \textbf{Cultural and Linguistic Variation:} While we support four languages, cultural factors influencing trust (e.g., collectivism vs. individualism, institutional trust levels, media literacy traditions) are not explicitly modeled. Moreover, LLM fluency varies across languages---models are typically more fluent in English than in lower-resource languages, potentially creating language-dependent ``fluency traps.'' Cross-cultural replication with language-stratified analysis is needed to validate whether susceptibility patterns generalize.
    \item \textbf{Longitudinal Effects:} Our study captures within-session learning but does not track retention or decay of detection skills over weeks or months. Inoculation research~\cite{roozenbeek2024prebunking} suggests effects may attenuate without booster exposures; understanding this decay curve is critical for designing sustainable intervention programs.
\end{itemize}

\textit{From Association to Intervention:} Future work must move beyond observing correlations to performing interventions. We propose using \ToolGen\ to generate text with specific ``causal flaws'' (e.g., impossible physics, logical contradictions, temporal inconsistencies). By measuring whether users can detect these causal errors better than stylistic errors, we can directly test the ``Level 2'' causal reasoning hypothesis of Chi et al.~\cite{chi2025unveiling}. This would reveal whether training users on causal consistency---rather than surface fluency---offers a more robust defense against AI-generated misinformation.

\textit{Closing the Trustworthy Cycle:} We envision a continuous improvement loop grounded in the framework of Han et al.~\cite{han2025trustworthy}:
\begin{enumerate}[leftmargin=*]
    \item \textbf{Detection:} Human feedback via \ToolEval\ identifies detection failures and susceptibility patterns.
    \item \textbf{Causal Discovery:} Analysis identifies the textual features ($M$) and user characteristics ($C$) driving these failures~\cite{liu2025causal}.
    \item \textbf{Model Improvement:} Discovered features inform ``safety classifiers'' or RLHF fine-tuning to reduce deceptive mimicry.
    \item \textbf{Re-evaluation:} Updated models are re-evaluated by humans, restarting the cycle.
\end{enumerate}

This \textit{Trustworthy Cycle}---integrating human perception, causal inference, and model refinement---provides a framework for developing trustworthy web ecosystems. Future work will incorporate additional foundation models (including Claude, Gemini, and Llama-3~\cite{meta2024llama3}) and investigate whether detection accuracy varies by content domain (e.g., political vs. scientific misinformation). Our ongoing research program includes detailed analysis of the AI-driven disinformation kill chain~\cite{loth2026collateraleffects} and reproduction of key findings from the disinformation literature~\cite{loth2026verification}.

\textbf{Open Science:} Both platforms are available as open-source projects. \ToolEval\ is accessible at \url{https://judgegpt.streamlit.app} (code: \url{https://github.com/aloth/JudgeGPT}), and \ToolGen\ at \url{https://github.com/aloth/RogueGPT}. We invite researchers to contribute to this ongoing investigation of human-AI interaction in the context of misinformation. As our study continues, we kindly invite experts to participate in our survey: \url{https://github.com/aloth/verification-crisis}.

\balance
\bibliographystyle{ACM-Reference-Format}
\bibliography{bibliography-arxiv-v3}

\clearpage

\appendix

\setcounter{page}{1}
\renewcommand{\thepage}{S\arabic{page}}
\renewcommand{\thefigure}{S\arabic{figure}}
\renewcommand{\thetable}{S\arabic{table}}
\setcounter{figure}{0}
\setcounter{table}{0}

\begin{center}
    \LARGE \textbf{Supplementary Materials}
\end{center}

This supplement documents the \ToolEval\ platform in detail.
We provide data schema descriptions and annotated user interface screenshots that illustrate the data collection methodology.
These materials facilitate replication and ensure transparency regarding the experimental design.

\section{Data Description}
\label{sec:data}

This study uses three main data collections: Participants, Fragments, and Results.
Each collection is stored in JSON format and provides granular attributes for in-depth analysis of participant attributes, fragment characteristics, and evaluation outcomes.

\subsection{Participants Collection}
The Participants collection provides demographic, linguistic, and contextual information about each participant.
Each participant is identified by a unique \texttt{ParticipantID}. Associated attributes include:
\begin{itemize}[nosep, leftmargin=*]
    \item \textit{Demographics:} \texttt{Age}, \texttt{Gender}, \texttt{EducationLevel}
    \item \textit{Attitudes:} \texttt{PoliticalView}, \texttt{FNewsExperience},\\ \texttt{NewspaperSubscription}
    \item \textit{Language:} \texttt{Language}, \texttt{IsNativeSpeaker}
    \item \textit{Context:} \texttt{ScreenResolution}, \texttt{IpLocation}, \texttt{UserAgent}
\end{itemize}
Political views are numerically encoded and mapped to categorical labels, as shown in Table~\ref{tab:politicalview}.
Familiarity with fake news (FNewsExperience) is encoded similarly, ranging from completely unfamiliar to completely familiar, with placeholder options where no choice is made (Table~\ref{tab:fnews}).
Additional contextual metadata, such as geolocation and browser information, provides environmental context for each participant’s judgments.

\begin{table}[!hb]
\centering
\caption{Political View Encoding}
\label{tab:politicalview}
\begin{tabular}{c l}
\toprule
Value & Label \\
\midrule
0.0 & Far Left \\
0.2 & Left \\
0.4 & Center-Left \\
0.5 & Placeholder (no choice) \\
0.6 & Center-Right \\
0.8 & Right \\
1.0 & Far Right \\
\bottomrule
\end{tabular}
\end{table}

\begin{table}[!hb]
\centering
\caption{Fake News Familiarity Encoding (FNewsExperience)}
\label{tab:fnews}
\begin{tabular}{c l}
\toprule
Value & Label \\
\midrule
0.0 & Completely unfamiliar \\
0.2 & Mostly unfamiliar \\
0.4 & Somewhat unfamiliar \\
0.5 & Placeholder (no choice) \\
0.6 & Somewhat familiar \\
0.8 & Mostly familiar \\
1.0 & Completely familiar \\
\bottomrule
\end{tabular}
\end{table}

\subsection{Fragments Collection}
The Fragments collection contains textual items (fragments) that participants are asked to evaluate.
Each fragment has attributes including \texttt{Language}, \texttt{Style}, \texttt{Format}, \texttt{SeedPhrase}, \texttt{FragmentID}, \texttt{Content}, \texttt{Origin} (human or machine), and \texttt{IsFake}.
Some fragments are styled to mimic certain news outlets or social media platforms, while others incorporate seed phrases that set topical or thematic content.
The \texttt{CreationDate} of each fragment is recorded, allowing temporal analysis.
This collection enables comparisons of different fragment properties (e.g., style, origin) and their effects on perceived authenticity.

\subsection{Results Collection}
The Results collection links participants to fragments and stores their assessment outcomes.
Each result record includes \texttt{ResultID}, \texttt{ParticipantID}, \texttt{FragmentID}, \texttt{HumanMachineScore}, \texttt{LegitFakeScore}, \texttt{TopicKnowledgeScore}, \texttt{Timestamp}, \texttt{TimeToAnswer}, \texttt{SessionCount}, \texttt{Origin}, and \texttt{IsFake}.
The HumanMachineScore, LegitFakeScore, and TopicKnowledgeScore are encoded on a 0.0 to 1.0 scale and map to interpretive labels, as shown in Tables~\ref{tab:humanmachine}, \ref{tab:legitfake}, and \ref{tab:topicknowledge}.
By combining Results with Participants and Fragments data, it is possible to analyze how participant attributes and fragment properties influence judgments of authenticity and how participants’ knowledge or familiarity impacts their decision-making.

\begin{table}[!hb]
\centering
\caption{Human vs. Machine Score Encoding (HumanMachineScore)}
\label{tab:humanmachine}
\begin{tabular}{c l}
\toprule
Value & Label \\
\midrule
0.0 & Definitely Human Generated \\
0.2 & Probably Human Generated \\
0.4 & Likely Human Generated \\
0.5 & Placeholder (no choice) \\
0.6 & Likely Machine Generated \\
0.8 & Probably Machine Generated \\
1.0 & Definitely Machine Generated \\
\bottomrule
\end{tabular}
\end{table}

\begin{table}[!hb]
\centering
\caption{Legit vs. Fake News Score Encoding (LegitFakeScore)}
\label{tab:legitfake}
\begin{tabular}{c l}
\toprule
Value & Label \\
\midrule
0.0 & Definitely Legit News \\
0.2 & Probably Legit News \\
0.4 & Likely Legit News \\
0.5 & Placeholder (no choice) \\
0.6 & Likely Fake News \\
0.8 & Probably Fake News \\
1.0 & Definitely Fake News \\
\bottomrule
\end{tabular}
\end{table}

\begin{table}[!hb]
\centering
\caption{Topic Knowledge Score Encoding (TopicKnowledgeScore)}
\label{tab:topicknowledge}
\begin{tabular}{c l}
\toprule
Value & Label \\
\midrule
0.0 & Not at all \\
0.2 & Slightly \\
0.4 & Somewhat \\
0.5 & Placeholder (no choice) \\
0.6 & Fairly well \\
0.8 & Very well \\
1.0 & Extremely well \\
\bottomrule
\end{tabular}
\end{table}

\section{JudgeGPT Interface Screenshots} 
\label{app:screenshots}

\begin{figure*}[!htbp] %
    \centering
    \includegraphics[width=0.65\textwidth]{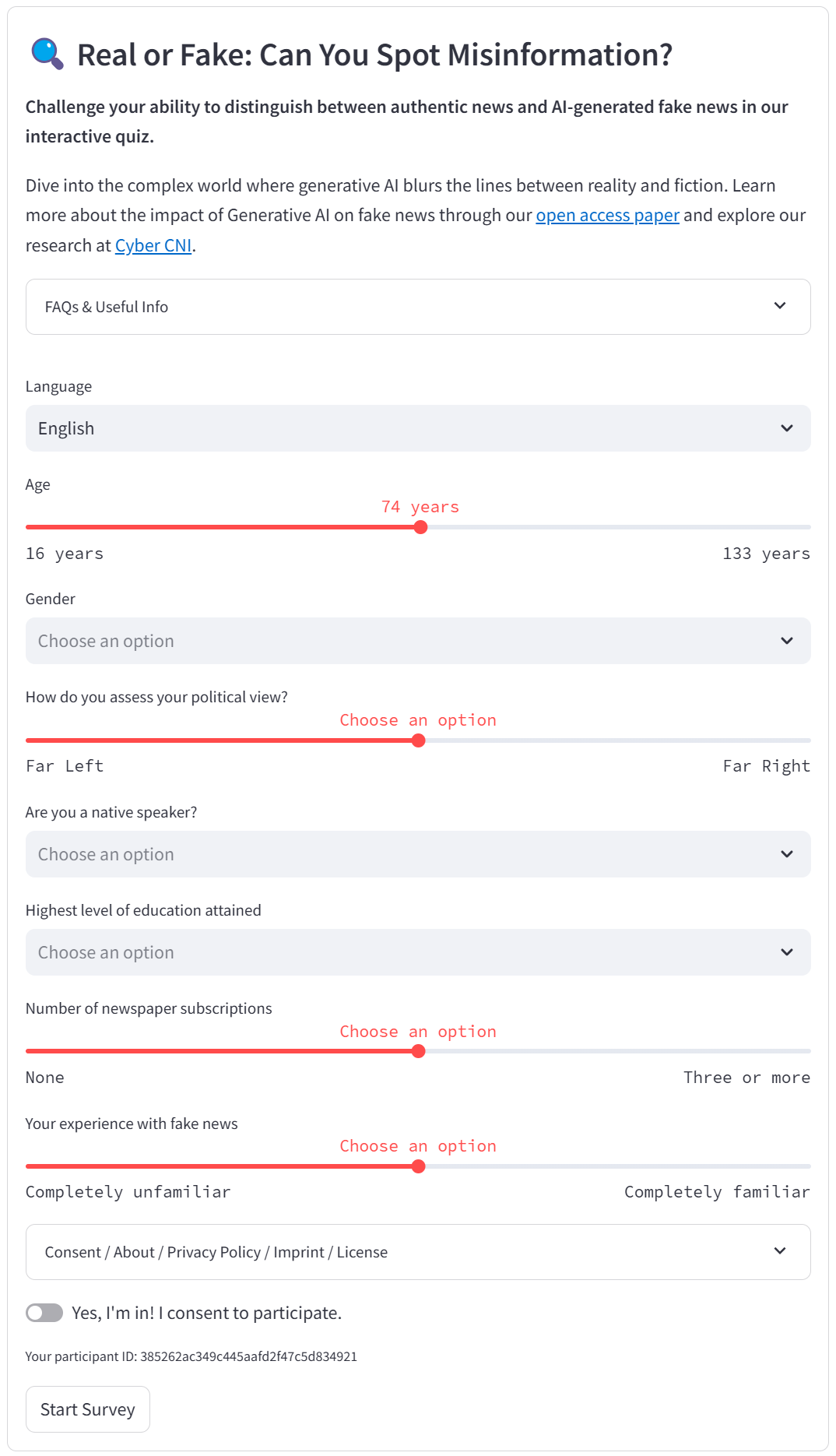} 
    \Description{A web interface screenshot titled ``Real or Fake: Can You Spot Misinformation?'' showing the JudgeGPT demographic questionnaire. The form includes: a language dropdown (set to English), an age slider ranging from 16 to 133 years (currently at 74), a gender dropdown, a political view slider from ``Far Left'' to ``Far Right,'' a native speaker question, highest education level dropdown, a newspaper subscriptions slider from ``None'' to ``Three or more,'' and a fake news experience slider from ``Completely unfamiliar'' to ``Completely familiar.'' At the bottom, an expandable section for Consent/About/Privacy Policy is shown, followed by a consent toggle (``Yes, I'm in! I consent to participate''), a unique participant ID, and a ``Start Survey'' button.}
    \caption{Screenshot of the \ToolEval\ Demographic Information Screen. This initial intake screen captures participant demographics including age, gender, education level, political orientation, and self-reported familiarity with fake news---variables central to our analysis (Section~\ref{sec:results}).}
    \label{fig:demographicscreen_appendix}
\end{figure*}

\begin{figure*}[!htbp] %
    \centering
    \includegraphics[width=0.72\textwidth]{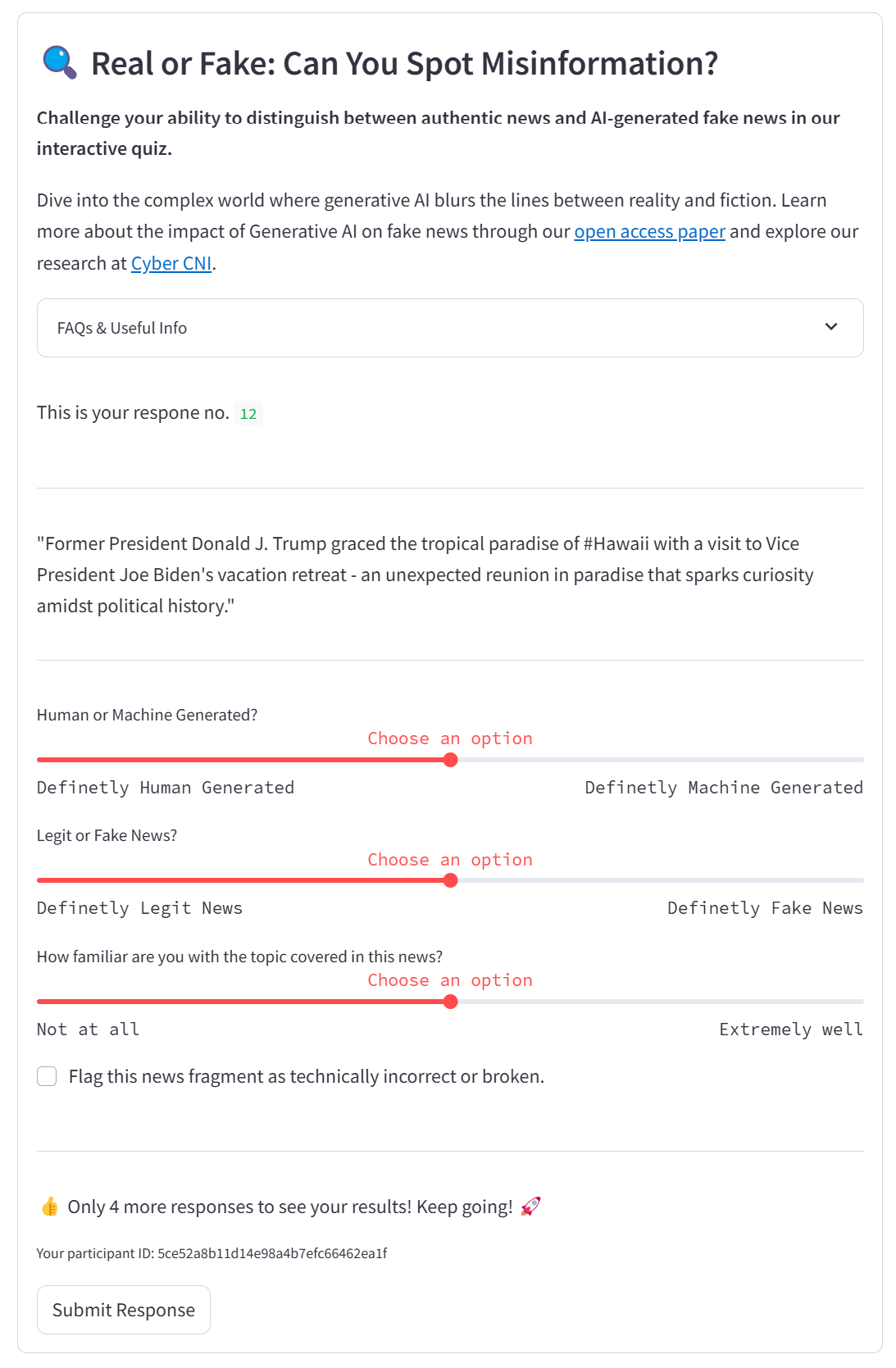} 
    \Description{A web interface screenshot showing the JudgeGPT evaluation screen, response number 12 of the survey. The main panel displays a news text fragment in quotation marks about a political topic. Below the text are three horizontal slider controls: the first labeled ``Human or Machine Generated?'' ranging from ``Definitely Human Generated'' to ``Definitely Machine Generated''; the second labeled ``Legit or Fake News?'' ranging from ``Definitely Legit News'' to ``Definitely Fake News''; and the third asking ``How familiar are you with the topic covered in this news?'' ranging from ``Not at all'' to ``Extremely well.'' A checkbox allows participants to flag technically incorrect or broken content. A progress message reads ``Only 4 more responses to see your results! Keep going!'' with emoji icons, followed by the participant ID and a ``Submit Response'' button.}
    \caption{Screenshot of the \ToolEval\ Survey Screen. Participants evaluate randomized news fragments using three continuous sliders: the \textit{HumanMachineScore} (ranging from ``Definitely Human Generated'' to ``Definitely Machine Generated''), the \textit{LegitFakeScore} (ranging from ``Definitely Legit News'' to ``Definitely Fake News''), and a topic familiarity rating. A flag option allows participants to report technically broken content. This multi-dimensional rating design enables independent measurement of source attribution accuracy and content credibility, as analyzed in Section~\ref{sec:results}.}
    \label{fig:surveyscreen_appendix}
\end{figure*}

The following figures document the \ToolEval\ web application interface, which was developed using Streamlit and deployed as an open-access platform. The platform implements a within-subjects design where each participant first completes a demographic intake form (Figure~\ref{fig:demographicscreen_appendix}) before evaluating a randomized sequence of news fragments (Figure~\ref{fig:surveyscreen_appendix}). This design enables us to capture both individual difference variables and learning effects across evaluation trials, consistent with Source Monitoring Framework methodology.

\end{document}